\documentclass[12pt, oneside, onecolumn]{article}

\usepackage{cite}
\usepackage[cmex10]{amsmath}
\usepackage{fixltx2e}
\usepackage{datetime}
\usepackage{graphicx}
\usepackage{stfloats}
\usepackage{color}
\usepackage[T1]{fontenc}
\usepackage[utf8]{inputenc}
\usepackage{authblk}
\usepackage{amssymb}
\usepackage{algorithm}
\usepackage{pstricks,epsfig}
\usepackage{rotating}
\usepackage{subfig}
\usepackage{tikz}
\usepackage{enumerate}
\usepackage{tabularx}
\usepackage{verbatim}
\usepackage{pgf}
\usepackage{url}
\usepackage[margin=1.75cm]{geometry}

\begin{document}

\begin{flushleft}
{\Large
\textbf{Bayesian-based deconvolution fluorescence microscopy using dynamically updated nonparametric nonstationary expectation estimates}
}
\\
$Alexander~Wong^{1,\ast}$,
$Xiao~Yu~Wang^{1}$, and
$Maud~Gorbet^{1}$
\\
\bf{1} Department of Systems Design Engineering, University of Waterloo, Ontario, Canada, N2L 3G1.
\\
$\ast$ E-mail: a28wong@uwaterloo.ca
\end{flushleft}


\textbf{Fluorescence microscopy is widely used for the study of biological specimens. Deconvolution can significantly improve the resolution and contrast of images produced using fluorescence microscopy; in particular, Bayesian-based methods have become very popular in deconvolution fluorescence microscopy.  An ongoing challenge with Bayesian-based methods is in dealing with the presence of noise in low SNR imaging conditions.  In this study, we present a Bayesian-based method for performing deconvolution using dynamically updated nonparametric nonstationary expectation estimates that can improve the fluorescence microscopy image quality in the presence of noise, without explicit use of spatial regularization.}

\section*{Introduction}

Fluorescence microscopy has become an essential tool for visualizing cellular structures and properties of biological specimens by using fluorescence as a means of identifying cellular components.  For example, using fluorescent live/dead stains as well as fluorescently-labeled particles or molecules (such as antibodies), researchers are able to characterize the effect of a treatment on cell population.  While many techniques allow for a quantitative assessment of cell phenotype, in the case of low cell numbers and when studying structure and/or viability of adherent cells, fluorescence microscopy remains the tool of choice.  Features of the biological sample and staining procedures may affect the quality of the acquired image.  Deconvolution fluorescence microscopy provides a way of overcoming the inherent limitations of the optical microscopy system by making use of the intrinsic properties of the system to increase image contrast and resolution after image acquisition.  This allows for improved visualization at subresolution scales.

A number of deconvolution methods have been previously introduced, and include nearest neighbor methods~\cite{Agard}, Wiener filtering~\cite{Erhardt}, iterative constrained regularization methods~\cite{Erhardt,Voort,Galatsanos,Tikhonov,Arigovindan}, and Bayesian-based methods~\cite{Richardson,Lucy,Trussell,Hunt,Hunt96,Orieux,Preibisch,Bertero1,Bertero2,Conchello,Dey,Remmele,Schaefer,Vicidomini,Zanella,Markham,Joshi}.  Bayesian-based deconvolution methods are statistical methods that model images, point spread functions (PSF), and imaging noise as probability distributions, and can be separated into two classes: maximum likelihood (ML) methods, and maximum a posteriori (MAP) methods.  While current Bayesian-based deconvolution methods account for inherent image noise characteristics, they are sensitive to low signal-to-noise ratio (SNR) imaging conditions where prior model parameter estimation is very challenging due to the presence of noise, thus leading to reduced deconvolved image quality.  A lot of focus on deconvolution fluorescence microscopy in dealing with the presence of noise in low SNR imaging conditions has revolved around the introduction of spatial regularization priors~\cite{Dey,Vicidomini} to impose spatial smoothness for suppressing noise characteristics.  Other strategies, such as ForWarD~\cite{forward}, have been proposed to deal with the presence of noise by first performing deconvolution followed by noise suppression~\cite{Donoho1,Donoho2,denoise1}.  In this study, we investigate and attempt to mitigate this issue from a different perspective by introducing an MAP method that performs deconvolution on fluorescence microscopy images using dynamically updated nonparametric nonstationary expectation estimates, which does not make explicit use of spatial regularization.

Here, we model the deconvolved fluorescence microscopy image ($f$) and the measured fluorescence microscopy image ($g$) as probability distributions.  In the proposed MAP method, the goal is to determine the most probable deconvolved image $\hat{f}$ given the measured image $g$, based on prior knowledge related to $f$.  The MAP problem can be formulated as:

\begin{equation}
	\hat{f} = {\rm argmax}_{f}~p\left(f | g\right),
\end{equation}

\noindent where $p\left( f | g \right)$ is the conditional probability of $f$ given $g$.  We model $g$ as Poisson distributed, and model $f$ as a nonstationary process with a nonstationary expectation ${E}(f_s)$ and a variance $\tau^2$, where $s$ denotes the pixel site location.  In previous MAP methods~\cite{Hunt}, the measured fluorescence microscopy image $g$ was used as an estimate of $E(f_s)$, based on the assumption that the measured image, having undergone the influence of the PSF, is similar to a nonstationary local average and thus representative of a nonstationary expectation ${E}(f_s)$.  However, this assumption may not be reliable in high noise situations under low SNR imaging conditions, as well as result in high-varying estimates depending on the inherent PSF.  One can obtain a more reliable estimate of $E(f_s)$ by computing local windowed averages of $g$; however, this method is also highly sensitive to the presence of high noise levels.  To tackle this issue, we introduce a dynamically updated nonparametric kernel-based estimate (based on the work by~\cite{Parzen,Rosenblatt}) of $E(f_s)$ that updates with each iteration.  Hence, we derive the iterative solution of the proposed MAP method (see \textbf{Methods} for full derivation), yielding:


\begin{equation}
	\hat{f}^{j+1}_s = \hat{f}^{j}_s\left[\frac{H_{-s} \bigotimes {g_s}}{H \bigotimes \hat{f}^{j}_s}\right]+\lambda \hat{f}^{j}_s \left(\frac{\sum_{i \in W_i}e^{-\beta(\hat{f}^{j}_i-\hat{f}^{j}_s)^2}\hat{f}^{j}_s}{\sum_{i \in W_i}e^{-\beta(\hat{f}^{j}_i-\hat{f}^{j}_s)^2}} - \hat{f}^{j}_s\right)
\label{iterationstep}
\end{equation}

\noindent where $\hat{f}^{j}_s$ denotes the deconvolved image at iteration $j$, $\bigotimes$ denotes convolution operator, $H$ denotes the point spread function (PSF), $\lambda$ and $\beta$ denote relaxation parameters, and $W_i$ denotes a window centered at location $i$.

Eq.~(\ref{iterationstep}) denotes an update step of the iteration solution.  The importance of introducing a dynamically updated nonparametric kernel-based estimate of $\hat{E}(f_s)$ that updates with each iteration is that it allows for a reliable estimate of $\hat{E}(f_s)$ under the influence of image noise.  Hence, the proposed MAP method results in a solution that remains stable under low SNR imaging conditions, leading to improved image quality of the deconvolved image $f$.  The proposed solution, to the best of our knowledge, is the first to consider to incorporate dynamically updated nonparametric nonstationary expectation estimates within a Bayesian theoretical framework for the purpose of deconvolution fluorescence microscopy, which allows for more stable solutions without the explicit use of spatial regularization.

\section*{Results}

To evaluate the proposed MAP method with dynamically updated nonparametric nonstationary expectation estimates (which we will refer to as MAP-D in figures), we first simulated a fluorescence microscopy data set with fluorescence-stained cell populations using a modified version of SIMCEP~\cite{Lehmussola}, a computational framework for simulating fluorescence microscopy images of fluorescence-stained cell populations (see Fig. 1b and Methods).  For comparison purposes, three other Bayesian-based methods were tested: Lucy-Richardson (LR) deconvolution~\cite{Lucy,Richardson}, Markov-Chain Monte-Carlo Wiener-Hunt (MCMC-WH) deconvolution~\cite{Orieux}, and Hunt MAP (MAP-Hunt) deconvolution method with Poisson likelihood and Gaussian image prior~\cite{Hunt}.  Note that LR, MAP-Hunt, and the proposed MAP-D methods are performed without the explicit use of spatial regularization to test the hypothesis that MAP-D can achieve improved fluorescence microscopy image quality in the presence of noise, without explicit use of spatial regularization.  The proposed MAP-D method (Fig. 1f) was able to achieve a significant increase in contrast and resolution when compared to the original fluorescence microscopy acquisition, as well as a noticeable SNR increase (Figs. 1f).  This is most evident by the fact that all 4 subcellular structures inside the cytoplasm of each cell are visible in the deconvolved image produced using the proposed MAP-D method when compared to the original fluorescence microscopy acquisition.  Compared to MAP-Hunt deconvolution~\cite{Hunt} (Fig. 1e), the proposed MAP-D method achieved greater increases in contrast and resolution, and noticeable SNR improvements.  This is particular relevant when studying bacteria binding to cells as the increase in contrast and resolution enables to accurately identify individual particles.  The proposed MAP-D method also provides added benefits when studying cell processes such as microvesicles and cell interactions with nanoparticles.

The proposed MAP-D method was then evaluated on fluorescence microscopy acquisitions of cells obtained from an ocular surface wash following sodium fluorescein instillation (see \textbf{Methods}).  Fig. 2a shows an acquisition of cell aggregates, some staining with fluorescein (green) while others stain red with the dead stain (ethidium bromide).  Fig. 2f shows a live/dead (calcein blue/ethidium bromide-red) acquisition of two corneal epithelial cells.  (Figs. 2a and 2f and \textbf{Methods}).  The proposed MAP-D method (Fig. 2e and 2j) was able to achieve a significant increase in contrast and resolution when compared to the original fluorescence microscopy acquisitions (Figs. 2a and 2f).  Compared to MAP-Hunt deconvolution~\cite{Hunt} (Fig. 2d and 2i), the proposed MAP-D method achieved greater increases in contrast and resolution, and moderate SNR improvements.    Based on a cross-sectional profile of an area of interest (Fig. 2k) and a background area (Fig. 2l) from Fig. 2a, we can observed that the proposed method improve contrast and resolution without introducing undesirable artifacts.  Additional deconvolution results from microscopy acquisitions in Fig. 3 further reinforce these observations.  In the original fluorescence microscopy acquisitions (Figs. 3a, 3f, 3k and 3p), due to low contrast levels, it was difficult to determine the nature of the cells present as the cell walls could not be clearly identified.  Deconvolution results achieved using LR, MAP-Hunt, and MCMC-WH resulted in a cell wall that is easier to detect but also introduced noticeable noise artifacts and thus reduced the overall image quality (Figs. 3b-3d, 3g-3i, 3l-3n and 3q-3s).   With the proposed MAP-D method, improved cell wall identification as well as improved contrast were obtained without increasing noise artifacts (Figs. 3f, 3k, 3o, 3t).

\section*{Discussion}

In this study, we introduce a Bayesian-based method for performing deconvolution using dynamically updated nonparametric nonstationary expectation estimates that can improve the fluorescence microscopy image quality.  The key factor to the improved performance of the proposed method compared to the other tested Bayesian-based deconvolution methods is the introduction of dynamically updated nonparametric expectation estimates into a Bayesian-based deconvolution framework, which greatly increases the stability of the solution in the presence of noise, without explicit use of spatial regularization.  Furthermore, by dynamically updating the expectation estimate at each iteration, the estimate of ${E}(f_s)$ becomes increasingly more reliable as the deconvoluted image estimate $\hat{f}_s$ becomes more accurate.  The only free parameters of the proposed MAP method are $\lambda$, $\beta$, $W$, and the number of iterations for deconvolution, which can be adjusted by the user to find a tradeoff between image quality and computational costs.

Analysis of ex vivo collected human cells present different challenges.  In the context of our study, due to the nature of the collection (gentle eye wash), the presence of debris can make cell identification difficult (as illustrated in Fig. 2a and 3a). Corneal epithelial cells collected from the ocular surface can also uptake various stains during the procedures; green from instillation of sodium fluorescein on the ocular surface, blue from calcein blue if the cell is alive and red from ethidium bromide if the cell is dead.  Ghost cells, cells without a nucleus, may also be present in the ocular wash. Current research aims at determining how corneal epithelial cells uptake fluorescein and how treatment, such as different combinations of lens cleaning solutions and contact lenses, can affect the ocular surface and cell shedding (the cells collected during the gentle eye wash).  Fluorescence microscopy allows to characterize these cells, however poor image quality can prevent efficient counting and investigation of the population of cells present in the ocular wash.  While confocal microscopy may be seen as an alternative, cost, cell transport, acquisition time and operator experience make confocal microscopy ill-suited for such investigations.  Other techniques have been used to collect and study these cells~\cite{collection1,collection2}; however, the image quality of fluorescein-stained cells was also a challenge, highlighting the need of a back-end approach to help resolve such issues as the cell collection technique is not able to provide significant improvements in the image quality of the cell samples.  This is likely due to the complexity of ex vivo cell collection, and the fragility of the cells studied which restricts processing protocols and times.  The ability to increase fluorescence image quality using the proposed MAP-D method so that cellular membranes and staining patterns can be better identified can further enhance the use and benefits of fluorescence microscopy.  Furthermore, increasing fluorescence image quality using the proposed MAP-D method could lead to improvements in tasks used for data analysis such as cell counting~\cite{cellcount} and data alignment~\cite{registration1,registration2}. As such, the proposed method can be valuable for improving fluorescence microscopy image quality to facilitate for the study of biological specimens, more specifically ex vivo collected samples where the presence of debris or other collection artifacts may exist and where significant cell processing is not an option due to low cell numbers and poor cell viability.  Future work involves the investigation of alternative nonparametric expectation estimation approaches, such as Monte Carlo estimation strategies~\cite{MCMC1,MCMC2,MCMC3,MCMC4}, to further improve fluorescence microscopy image quality in the presence of noise.  

\section*{Methods}
~\\
\textbf{Derivations}.
~\\
A full derivation of the iterative solution of the proposed MAP method can be described as follows.  Let $S$ be a set of pixel locations into a discrete lattice $\mathcal L$ and $s \in S$ be a pixel location in $\mathcal L$.  Let $F=\{F_s|s \in S\}$ and $G=\{G_s|s \in S\}$ be random fields on $S$, with $F_s$ and $G_s$ taking on values representing the deconvolved fluorescence microscopy image and measured fluorescence microscopy image at pixel $s$ respectively.  Let $f=\{f_s|s \in S\}$ and $g=\{g_s|s \in S\}$ be realizations of $F$ and $G$ respectively.  Let $H$ denote the point-spread function.  The goal is to estimate $f$ given $g$, based on prior knowledge related to $f$.

\begin{equation}
	\hat{f} = {\rm argmax}_{f}~p\left(f | g\right),
\end{equation}

\noindent this is equivalent to:

\begin{equation}
	\hat{f} = {\rm argmax}_{f}~p\left(g | f\right)p(f),
\label{MAP2}
\end{equation}

\noindent where $p(g | f)$ is the likelihood and $p(f)$ is the prior.  In the case of fluorescence microscopy, image noise is primarily related to quantum photon emission and as such can express $p(g|f)$ as follows:

\begin{equation}
	p\left(g | f\right) = \prod_{s \in S} \frac{\left(H \bigotimes f_s\right)^{{g_s}}e^{-(H \bigotimes f_s)}}{{g_s}!}
\end{equation}

\noindent where $\bigotimes$ denotes the convolution operator.  As a prior model $p(f)$, we employ a Gaussian prior model with a nonstationary expectation ${E}(f_s)$ and a variance $\tau^2$:

\begin{equation}
	p\left(f\right) = \prod_{s \in S} e^{-\frac{\left(f_s-{E}(f_s)\right)^2}{2 \tau^2}},
\end{equation}

\noindent Therefore, Eq.~(\ref{MAP2}) can be reformulated as:
\begin{equation}
	\hat{f} = {\rm argmax}_{f}~\left(\prod_{s \in S} \frac{\left(H \bigotimes f_s\right)^{{g_s}}e^{-(H \bigotimes f_s)}}{{g_s}!}e^{-\frac{\left(f_s-{E}(f_s)\right)^2}{2 \tau^2}}\right).
\label{MAP3}
\end{equation}
\noindent This is equivalent to minimizing the negative logarithm of $p\left(g | f\right)p(f)$, given us:
\begin{equation}
	\hat{f} = {\rm argmin}_{f}~\int \left[\left(H \bigotimes f_s\right)-{g_s} \log \left(H \bigotimes f_s\right) + \log \left( {g_s}! \right)+\frac{1}{2 \tau^2}\left(f_s-{E}(f_s)\right)^2\right]ds.
\label{MAP4}
\end{equation}
\noindent To find the solution that minimizes Eq.~(\ref{MAP4}), we take the derivative and set it to zero:
\begin{equation}
H_{-s} \bigotimes \left[1 - \frac{{g_s}}{\left(H \bigotimes f_s\right)}\right]+\lambda \left(f_s-{E}(f_s)\right) = 0
\label{MAP4a}
\end{equation}
\noindent where $\lambda$ is a relaxation parameter.  Simplifying further:
\begin{equation}
	 H_{-s} \bigotimes 1 -\left[\frac{H_{-s} \bigotimes {g_s}}{H \bigotimes {f}_s}\right]-\lambda \left({E}(f_s) - {f}_s\right) = 0
\label{MAP4b}
\end{equation}
\begin{equation}
	 \left[\frac{H_{-s} \bigotimes {g_s}}{H \bigotimes {f}_s}\right]+\lambda \left({E}(f_s) - {f}_s\right) = 1.
\label{MAP5}
\end{equation}

To establish the iterative solution, let us assume that $\frac{\hat{f}^{j+1}_s}{\hat{f}^{j}_s}=1$ at convergence, where $\hat{f}^{j}_s$ denotes the deconvolved image at iteration $j$.  Therefore,

\begin{equation}
	 \left[\frac{H_{-s} \bigotimes {g_s}}{H \bigotimes \hat{f}^{j}_s}\right]+\lambda \left({E}(f_s) - \hat{f}^{j}_s\right) = \frac{\hat{f}^{j+1}_s}{\hat{f}^{j}_s}
\end{equation}

\begin{equation}
	\hat{f}^{j+1}_s = \hat{f}^{j}_s\left[\frac{H_{-s} \bigotimes {g_s}}{H \bigotimes \hat{f}^{j}_s}\right]+\lambda \hat{f}^{j}_s \left({E}(f_s) - \hat{f}^{j}_s\right)
\label{iterationstep1}
\end{equation}

Since ${E}(f_s)$ is unknown, we instead use an estimate $\hat{E}(f_s)$, giving us:
  \begin{equation}
	\hat{f}^{j+1}_s = \hat{f}^{j}_s\left[\frac{H_{-s} \bigotimes {g_s}}{H \bigotimes \hat{f}^{j}_s}\right]+\lambda \hat{f}^{j}_s \left(\hat{E}(f_s) - \hat{f}^{j}_s\right)
\label{iterationstep2}
\end{equation}

  Contrary to existing methods~\cite{Hunt}, which employ the measured fluorescence microscopy image $g$ as an estimate of $\hat{E}(f_s)$, we introduce a dynamically updated nonparametric kernel-based estimate of $\hat{E}(f_s)$ that updates with each iteration for a more reliable estimate of $\hat{E}(f_s)$ under the influence of noise, particularly low-SNR scenarios, which can be expressed by:
    \begin{equation}
    \hat{E}(f_s) = \frac{\sum_{i \in W_s}K{(\hat{f}^{j}_i-\hat{f}^{j}_s)}\hat{f}^{j}_s}{\sum_{i \in W_s}K{(\hat{f}^{j}_i-\hat{f}^{j}_s)}},
    \label{kernelmean}
    \end{equation}

    \noindent where $W_s$ denotes a window centered at location $s$ and $K$ is a kernel function.  Plugging Eq.~(\ref{kernelmean}) into Eq.~(\ref{iterationstep2}), with the following Gaussian kernel function:
     \begin{equation}
     K(f_i-f_s)=e^{-\beta(f_i-f_s)^2}
      \end{equation}
      \noindent where $\beta$ is a relaxation parameter, gives us the final iterative solution of the proposed MAP method, yielding:

\begin{equation}
	\hat{f}^{j+1}_s = \hat{f}^{j}_s\left[\frac{H_{-s} \bigotimes {g_s}}{H \bigotimes \hat{f}^{j}_s}\right]+\lambda \hat{f}^{j}_s \left(\frac{\sum_{i \in W_i}e^{-\beta(\hat{f}^{j}_i-\hat{f}^{j}_s)^2}\hat{f}^{j}_s}{\sum_{i \in W_i}e^{-\beta(\hat{f}^{j}_i-\hat{f}^{j}_s)^2}} - \hat{f}^{j}_s\right)
\label{iterationstepp}
\end{equation}
~\\
\textbf{Imaging apparatus}.
~\\
For this study, the Nikon fluorescence microscope Eclipse, TE2000-S, with a triple band fluorescence filter module (DAPI/FITC/TRITC) was used (Fig. 3k) to obtain fluorescence microscopy acquisitions of cells obtained from an ocular surface wash following sodium fluorescein instillation. The microscope was equipped with a motorized XY stage and a microcontroller from Applied Scientific Instrument.  A DS-Fi1 digital camera and its controller as well as the NIS-Elements software for multi-dimensional acquisition were purchased from Nikon. Image acquisition was performed using a 20x objective (resulting in a 200x magnification when accounting for magnification of the eye piece).  Before deconvolution on fluorescence microscopy imaging data can take place with the proposed MAP method, it is first necessary to estimate the PSF of the microscope.  In this study, the Nikon fluorescence microscope Eclipse TE2000-S was characterized based on a number of image acquisitions of point source targets to estimate the PSF of the microscope.  In the situation where the PSF of the microscope is not known, PSFs obtained using simulations or estimated using other means may also be used in the proposed MAP deconvolution method.\\
~\\
\textbf{Simulation of fluorescence microscopy data sets}.
~\\
To evaluate the proposed MAP method, we simulated a ground-truth fluorescence microscopy data set with fluorescence-stained cell populations (Fig. 1a) and a convolved and noisy fluorescence microscopy data set from the same cell population (Fig. 1b) using a modified version of SIMCEP~\cite{Lehmussola}, a computational framework for simulating fluorescence microscopy images of fluorescence-stained cell populations.  The only modification to SIMCEP is the introduction of an additional Poisson process to generate pixel intensities in the SIMCEP framework.  We configure SIMCEP to generate cell populations with visible cytoplasm, nuclei, and a total of 4 subcellular structures inside the cytoplasm of each cell, and allowing for overlapping cells, with energy of autofluorescence and variance of CCD detector noise at 0.05 and 0.001, respectively.\\
~\\
\textbf{Sample preparation}.
~\\
Sample preparation was performed for this study in the following manner.  Following sodium fluorescein instillation, cells from the ocular surface were collected by a gentle eye wash using the Ocular Surface Cell Collection Apparatus~\cite{Peterson}.  Cells were centrifuged and then stained using the live/dead stain (calcein blue/ethidium bromide).  Following a 45 minute incubation in the dark at room temperature, cells were imaged immediately using a Nikon Eclipse fluorescence microscope.  Some cells uptook sodium fluorescein and appeared green; live cells stained with Calcein blue and their nucleus appeared blue while dead cells stained with ethidium bromide and were characterized by a red nucleus.\\
~\\
\textbf{Implementation details}.
~\\
The proposed MAP method (referred to here as MAP-D) is implemented in MATLAB (The MathWorks, Inc.), with the nonparametric kernel-based estimate of $\hat{E}(f(s))$ implemented in C++ and compiled a dynamically linked MATLAB Executable (MEX) for to improve computational speed.  The only free parameters of the proposed MAP-D method are $\lambda$, $\beta$, $W$, and the number of iterations for deconvolution, which can be adjusted by the user to find a tradeoff between image quality and computational costs.  For the experiments using fluorescence microscopy acquisitions, the parameters $\lambda$, $\beta$, $W$, and the number of iterations for deconvolution for MAP-D are set to 0.2, 625, a 81-sample window, and 50, respectively.  For Lucy-Richardson (LR) deconvolution~\cite{Lucy,Richardson}, the only free parameter was the number of iterations, and that was set to 50.   For the Hunt MAP (MAP-Hunt) deconvolution method with Poisson likelihood and Gaussian image prior~\cite{Hunt}, the free parameters are $\lambda$ and the number of iterations, and these were set to 0.2 and 50, respectively.  All parameters are chosen to be consistent across the LR, MAP-Hunt, and MAP-D methods (e.g., all shared parameters are the same) and provide strong image contrast and resolution improvements across tested acquisitions.  Finally, for Markov-Chain Monte-Carlo Wiener-Hunt (MCMC-WH) deconvolution~\cite{Orieux}, the code was provided in MATLAB by the authors of~\cite{Orieux}, and the free parameters are the number of iterations for burn-in and the maximum number of iterations, and these were set to 30 and 150, respectively, based on author-provided information.  The PSF of the microscope obtained as discussed above is used as an input parameter for all tested deconvolution methods.  For these configurations, the implementations of MAP-D, LR, MAP-Hunt, and MCMC-WH can process 800 $\times$ 800 three-channel fluorescence microscopy acquisitions in $\sim$45s, $\sim$39s, $\sim$40s, and $\sim$79s, respectively, on an Intel(R) Core(TM) i5-3317U CPU at 1.70GHz CPU with 4Gb RAM.



\section*{Acknowledgments}

This work was supported by the Natural Sciences and Engineering Research Council of Canada, Canada Research Chairs Program, and the Ontario Ministry of Research and Innovation.

\section*{Author contributions}

A.W. conceived and designed the method.  A.W. and X.W. worked on formulation and derivation of method solution.  M.G. performed the data collection.  A.W. performed the data processing.  M.G. performed the data analysis. All authors contributed to writing the paper and to the editing of the paper.

\section*{Competing Financial Interests}

The authors declare no competing financial interests.

\pagebreak
\section*{Figure Legends}

\begin{figure*}[!b]
	\centering
    \includegraphics[width=1\linewidth]{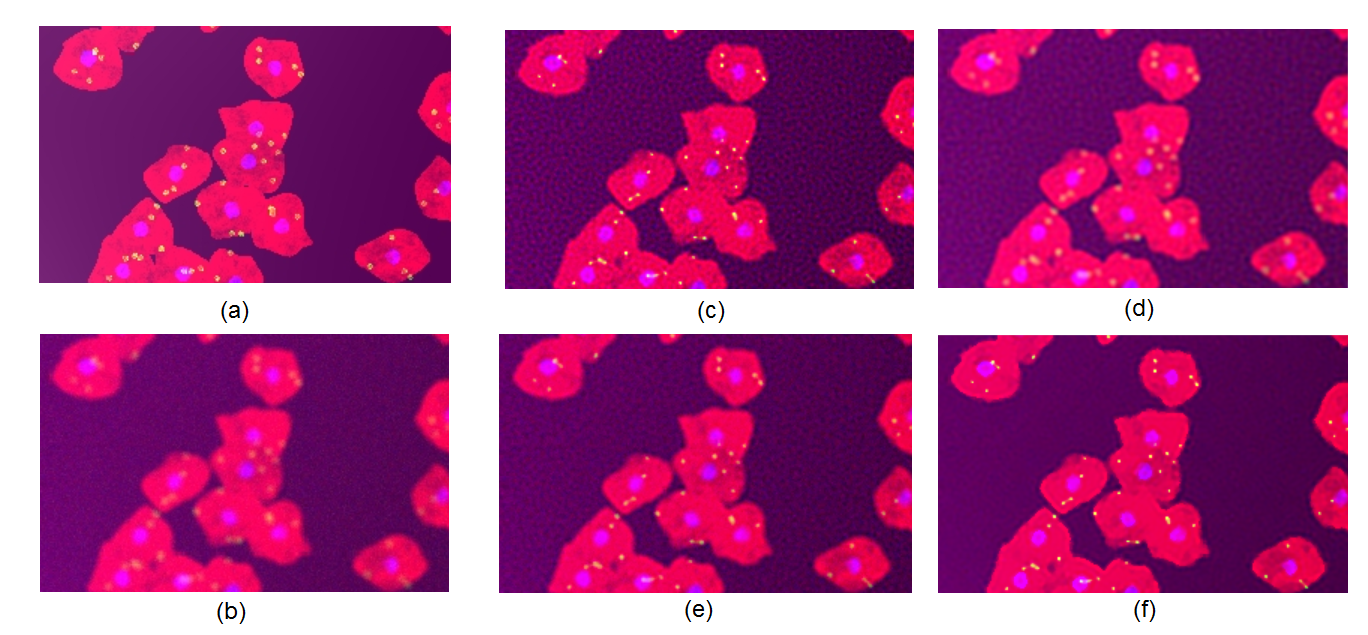}
	\caption{Application to simulated fluorescence microscopy data of fluorescence-stained cell population. \textbf{(a)} ground-truth simulated fluorescence microscopy imaging data and \textbf{(b)} simulated measured fluorescence microscopy imaging data.  \textbf{(c)-(f)} deconvolution results for \textbf{b} (\textbf{(c)}: LR, \textbf{(d)}: MCMC-WH, \textbf{(e)}: MAP-Hunt, \textbf{(f)}: MAP-D.  Contrast is significantly improved in MAP-D when compared to \textbf{b}, along with noticeable SNR increases.  }
	\label{fig1}
\end{figure*}

\begin{figure*}[!b]
	\centering
    \includegraphics[width=1\linewidth]{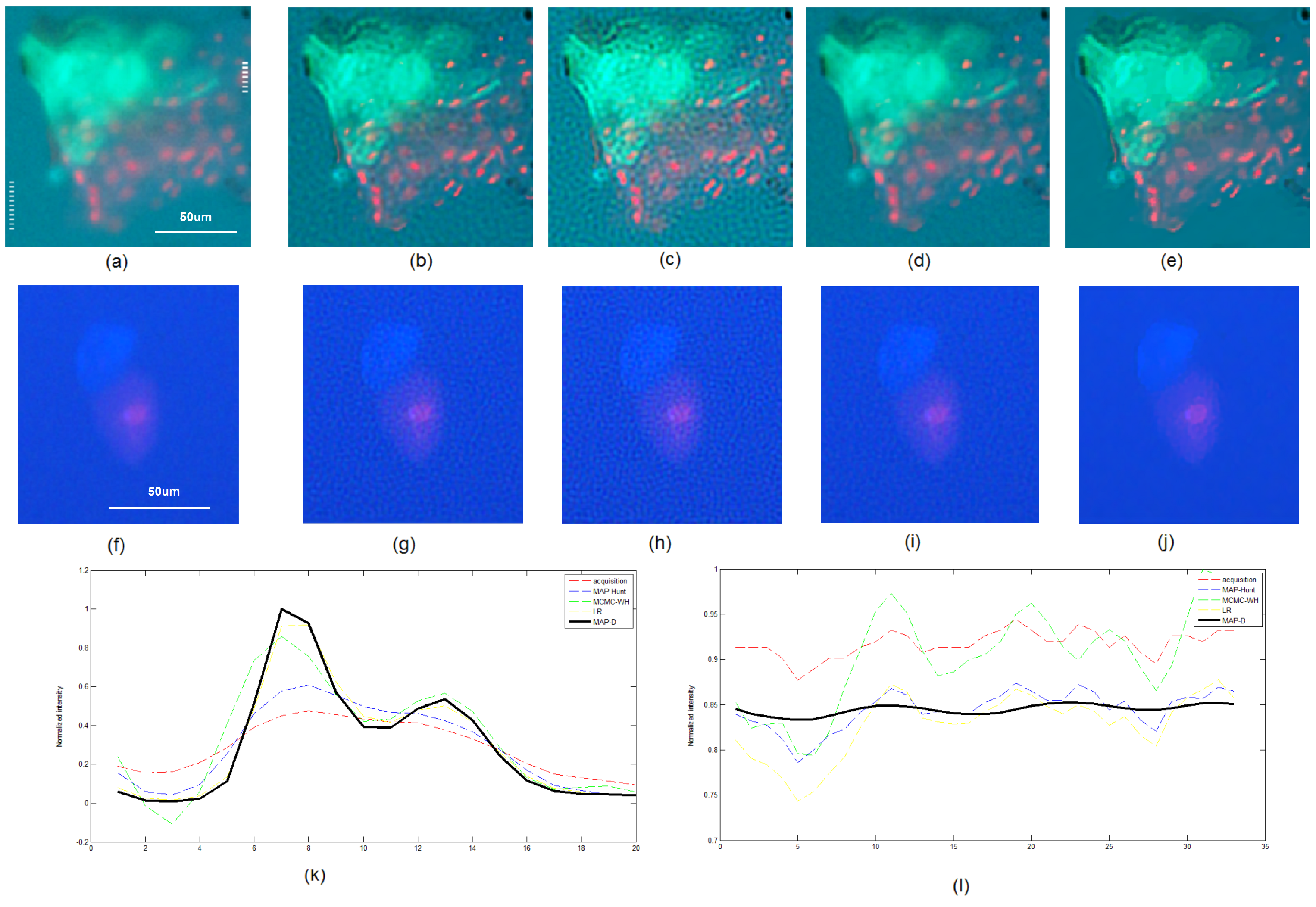}
	\caption{Application to fluorescence microscopy data of biological specimens, using fluorescence microscopy imaging of ocular surface wash.  \textbf{(a)} Cell aggregates can be observed, some staining with fluorescein (green) while others stain red with the dead stain (ethidium bromide).  \textbf{(f)} live/dead (calcein blue/ethidium bromide-red) fluorescent image of two corneal epithelial cells. The left and right dotted white lines found in \textbf{a} mark the plots shown in \textbf{k} and \textbf{l}, respectively. The top row corresponds to deconvolution results for \textbf{a} (\textbf{(b)}: LR, \textbf{(c)}: MCMC-WH, \textbf{(d)}: MAP-Hunt, \textbf{(e)}: MAP-D.  Contrast and resolution is significantly improved in MAP-D when compared to \textbf{a}, along with noticeable SNR increases when compared to \textbf{b-d}.  The bottom row corresponds to deconvolution results for \textbf{f} (\textbf{(g)}: LR, \textbf{(h)}: MCMC-WH, \textbf{(i)}: MAP-Hunt, \textbf{(j)}: MAP-D.  Similarly, contrast and resolution is significantly improved in MAP-D when compared to \textbf{f}, along with noticeable SNR increases when compared to \textbf{g-j}.  \textbf{(k)} Intensity-normalized line plot (red channel) through an area of interest in \textbf{a}.  Contrast is significantly enhanced in MAP-D when compared to the original acquisition.  \textbf{(l)} Intensity-normalized line plot (blue channel) through an background area in \textbf{a}.  SNR is noticeably improved in MAP-D when compared the other tested methods.}
	\label{fig1}
\end{figure*}

\begin{figure*}[!b]
	\centering
    \includegraphics[width=1\linewidth]{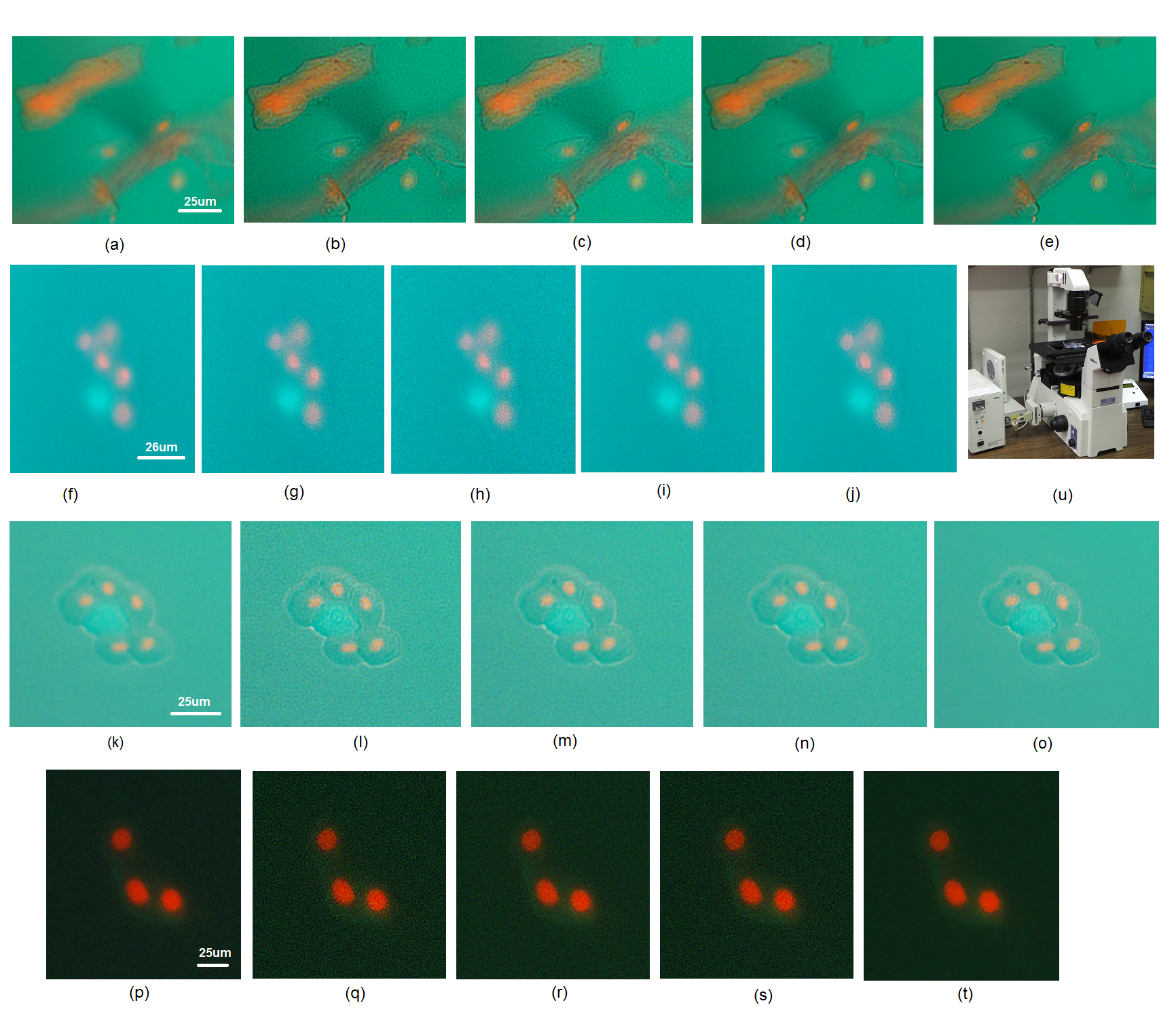}
	\caption{Application to fluorescence microscopy data of biological specimens. \textbf{(a)} fluorescence microscopy imaging data of an ocular surface wash following sodium fluorescein instillation.  Debris and dead corneal epithelial cells can be observed. The deconvolution results for \textbf{a} (\textbf{(b)}: LR, \textbf{(c)}: MCMC-WH, \textbf{(d)}: MAP-Hunt, \textbf{(e)}: MAP-D).  Contrast and resolution is significantly improved in MAP-D when compared to \textbf{a}, along with noticeable SNR increases when compared to \textbf{b-d}.  \textbf{(f) and (k)} fluorescence microscopy imaging data of an ocular surface wash following sodium fluorescein instillation. A potential fluorescein-stained cell (green staining) surrounded by dead corneal epithelial cells are somewhat visible. The deconvolution results for \textbf{f,k} (\textbf{(g,l)}: LR, \textbf{(h,m)}: MCMC-WH, \textbf{(i,n)}: MAP-Hunt, \textbf{(j,o)}: MAP-D).  Contrast and resolution is significantly improved in MAP-D when compared to \textbf{f} and \textbf{k}, along with noticeable SNR increases when compared to \textbf{g-i} and \textbf{l-n}, respectively.  \textbf{(p)} fluorescence microscopy imaging data of an ocular surface wash following sodium fluorescein instillation:  three potential fluorescein-stained cell (green intra-cellular staining) and each nucleus stained with ethiudium bromide (red) indicating cell death. The deconvolution results for \textbf{p} (\textbf{(q)}: LR, \textbf{(r)}: MCMC-WH, \textbf{(s)}: MAP-Hunt, \textbf{(t)}: MAP-D). The contrast of the cell wall is improved in MAP-D compared to \textbf{p} with SNR increases compared to \textbf{q-s}. \textbf{(u)} imaging apparatus for obtaining fluorescence microscopy acquisitions of cells obtained from an ocular surface wash following sodium fluorescein instillation.}
	\label{fig3}
\end{figure*}

\end{document}